\documentclass[aps,prb,10pt,twocolumn,superscriptaddress]{revtex4-1}

\usepackage{amsmath, amssymb}    					
\usepackage{graphicx}   							
\usepackage{xcolor}     							
\usepackage{hyperref}   							
\hypersetup{colorlinks=true,allcolors=blue}			
\usepackage{textcomp}								

\newcommand{\ddt}[0]{\frac{\partial}{\partial t}}
\renewcommand{\t}[1]{\textrm{#1}}
\newcommand{\nn}[0]{\nonumber\\}

\newcommand{\mbf}[1]{\mathbf{#1}}

\newcommand{\K}[0]{\mathbf{K}}

\newcommand{\up}[0]{\uparrow}
\newcommand{\down}[0]{\downarrow}
\newcommand{\Jsd}[0]{J_{sd}}
\newcommand{\Jpd}[0]{J_{pd}}

\newcommand{\nMn}[0]{n_\t{Mn}}
\newcommand{\omMn}{\omega_{\t{Mn}}}
\newcommand{\ome}{\omega_{\t{e}}}

\newcommand{\gel}{g_\t{e}}

\newcommand{\gMn}{g_\t{Mn}}

\newcommand{\etae}{\eta_{\t{e}}}
\newcommand{\etah}{\eta_{\t{h}}}
\renewcommand{\Im}{\textrm{Im}}
\renewcommand{\Re}{\textrm{Re}}
\newcommand{\ud}{{\uparrow/\downarrow}}
\newcommand{\du}{{\downarrow/\uparrow}}

\newcommand{\omsf}[0]{\omega_\mathrm{sf}}
\newcommand{\ph}[1]{\phantom{#1}}

\begin{document}

\title{Role of excited states in the dynamics of excitons and their spins in diluted magnetic semiconductors}
\author{F. Ungar}
\affiliation{Theoretische Physik III, Universit\"at Bayreuth, 95440 Bayreuth, Germany}
\author{M. Cygorek}
\affiliation{Department of Physics, University of Ottawa, Ottawa, Ontario, Canada K1N 6N5}
\author{V. M. Axt}
\affiliation{Theoretische Physik III, Universit\"at Bayreuth, 95440 Bayreuth, Germany}

\begin{abstract}

We theoretically investigate the impact of excited states on the dynamics of the exciton ground state in diluted magnetic semiconductor quantum wells.
Exploiting the giant Zeeman shift in these materials, an external magnetic field is used to bring transitions between the exciton ground state and excited states close to resonance.
It turns out that, when treating the exciton dynamics in terms of a quantum kinetic theory beyond the Markov approximation, higher exciton states are populated already well below the critical magnetic field required to bring the exciton ground state in resonance to an excited state.
This behavior is explained by exciton-impurity correlations that can bridge energy differences on the order of a few meV and require a quantum kinetic description beyond the independent-particle picture.
Of particular interest is the significant spin transfer toward states on the optically dark $2p$ exciton parabola which are protected against radiative decay.

\end{abstract}

\maketitle

\section{Introduction}
\label{sec:Introduction}

Ever since the seminal works by Frenkel \cite{Frenkel_On-the_1931} and Wannier \cite{Wannier_The-Structure_1937}, excitons in semiconductors have continued to attract attention and are nowadays routinely used in the optical characterization of materials \cite{Koch_Semiconductor-excitons_2006}.
Most notably, exciton states with very high principal quantum numbers have recently been experimentally observed in cuprous oxide \cite{Kazimierczuk_Giant-Rydberg_2014}, and exciton binding energies of several hundred meV have been found in transition-metal dichalcogenide (TMD) monolayers \cite{Ugeda_Giant-bandgap_2014, Chernikov_Exciton-Binding_2014, Brem_Exciton-Relaxation_2018}.
Although exciton binding energies are much smaller in standard bulk semiconductors, they can be significantly enhanced up to several tens of meV in semiconductor nanostructures such as quantum wells and wires \cite{Bastard_Wave-mechanics_1996}.
Here, we theoretically study excitons in diluted magnetic semiconductors (DMSs), where a small number of impurity ions with a large magnetic moment such as manganese is incorporated in the crystal lattice \cite{Furdyna_Diluted-magnetic_1988, Furdyna_Semiconductor-and_1988, Kossut_Introduction-to_2010, Dietl_Dilute-ferromagnetic_2014}.
To obtain sufficiently high binding energies, we consider Zn$_{1-x}$Mn$_{x}$Se quantum wells.
Being a II-VI semiconductor, ZnSe also allows for an isoelectronic incorporation of Mn impurities without the generation of excess carriers.

It is well known that correlations can play a decisive role in the magnitude of material properties such as band gaps and quasiparticle energies \cite{Hybertsen_Electron-correlation_1986, Ugeda_Giant-bandgap_2014} or their optical properties \cite{Kner_Coherence-of_1998, Axt_How-Correlated_2000, Shumway_Correlation-versus_2001, Chemla_Many-body_2001}.
DMSs are especially prominent materials in that regard since they are known to display strong correlation effects \cite{Ohno_Making-Nonmagnetic_1998, DiMarco_Electron-correlations_2013, Dietl_Dilute-ferromagnetic_2014, Ungar_Many-body_2018}, which is in part due to the large coupling constants found in the carrier-impurity exchange interaction.
This mechanism, which typically dominates the spin dynamics in DMSs \cite{Jiang_Electron-spin_2009, Smits_Excitonic-enhancement_2004}, describes a spin-flip scattering of carriers at the localized impurities.
Apart from its impact on the dynamics, the carrier-impurity exchange interaction also causes the giant Zeeman effect that significantly enhances the Zeeman splitting in an external magnetic field \cite{Dietl_Dilute-ferromagnetic_2014}.
By applying a magnetic field, one can thus bring the exciton ground state with appropriate spin into or close to resonance with an excited exciton state so transitions between them can occur easily.

In this paper, we study the impact of excited states on the dynamics of the optically excited exciton ground state in terms of both its occupation and its spin.
The simulations are performed for a system with a magnetic field that is tuned such that the $1s$ heavy-hole (hh) exciton is energetically close to one of the $2p$ states.
Since correlations are important in DMSs, as mentioned previously, we describe the exciton dynamics in terms of a quantum kinetic theory (QKT) which explicitly captures exciton-impurity correlations beyond the mean-field level \cite{Ungar_Quantum-kinetic_2017}.

It turns out that correlations significantly impact the dynamics, causing a sizable population of optically dark $2p$ excitons already well below the magnetic field required for a resonance of the $2p$ state with the ground state.
In contrast, a standard Markovian theory (MT), where all correlation effects are neglected such that excitons are effectively described as independent entities, yields only a finite occupation of the $2p$ state above this critical field.
This makes clear that the large correlation energies found in DMS quantum wells \cite{Ungar_Many-body_2018} allow for a bridging of otherwise still off-resonant transitions to higher exciton states.
It is worthwhile to note that the $2p$ excitons populated in this way cannot be directly addressed by optical excitation since they are dark.
The mechanisms discussed in this paper thus allow a transfer to states where the carrier spins are protected against radiative decay even after a relaxation towards the subband minimum.
Our analysis also reveals that the occupation of excited states is reflected in the spin dynamics of the exciton ground state, which becomes accelerated compared with a simulation where only the ground state is accounted for.

\section{Theory}
\label{sec:Theory}

First, we discuss the Hamiltonian used to model the exciton dynamics in DMSs and provide the equation of motion for the time-dependent occupation and spin of any given exciton state.
We first focus on results obtained by a recently developed quantum kinetic description of exciton spins in DMSs quantum wells \cite{Ungar_Quantum-kinetic_2017}, which explicitly takes exciton-impurity correlations into account, and we discuss its Markov limit in the following section.
In this limit, all correlations and thereby caused memory effects are disregarded to obtain a rate-type description.

\subsection{Quantum kinetic model}
\label{subsec:Quantum-kinetic-model}

We consider a II-VI DMS quantum well at a fixed temperature of $1\,$K which is optically excited with a short laser pulse.
In its ground state, such a semiconductor compound has a completely filled valence band and an empty conduction band.
If, additionally, an external magnetic field is applied along the growth direction, the Hamiltonian comprises the following parts \cite{Ungar_Quantum-kinetic_2017}:
\begin{align}
\label{eq:Hamiltonian}
H =&\; H_0^\t{e} + H_0^\t{h} + H_\t{conf} + H_\t{C} + H_\t{Z}^\t{e} + H_\t{Z}^\t{h} + H_\t{Z}^\t{Mn}
	\nn
	& + H_\t{lm} + H_{sd} + H_{pd} + H_\t{nm}^\t{e} + H_\t{nm}^\t{h}.
\end{align}
The kinetic energies of electrons and holes given by $H_0^\t{e}$ and $H_0^\t{h}$, respectively, together with the confinement $H_\t{conf}$ due to the quantum well and the Coulomb interaction given by $H_\t{C}$ define the exciton problem.
Its eigenfunctions are the exciton wave functions, labeled by their center-of-mass momentum $K$ and a discrete exciton quantum number $x \in \{1s, 2s, 2p, ...\}$ similar to that of the hydrogen problem in two dimensions, and its eigenvalues provide the corresponding energies.

The external magnetic field causes a Zeeman shift of electrons and holes given by $H_\t{Z}^\t{e}$ and $H_\t{Z}^\t{h}$, respectively, but also similarly affects the magnetic impurities via $H_\t{Z}^\t{Mn}$.
Furthermore, the interaction of the system with the laser pulse is contained in the light-matter interaction term $H_\t{lm}$, for which we use the usual dipole approximation \cite{Rossi_Theory-of_2002}.

The typically most important interaction in DMSs is the Kondo-type carrier-impurity exchange interaction \cite{Kossut_Introduction-to_2010, Thurn_Quantum-kinetic_2012, Dietl_Dilute-ferromagnetic_2014, Ungar_Quantum-kinetic_2017} given by $H_{sd}$ and $H_{pd}$.
These terms describe the spin-flip scattering of $s$-like conduction band electrons and $p$-like valence band holes with the localized electrons in the $d$ shell of an impurity ion, such as manganese.
Apart from a spin-flip scattering, impurities in general also cause nonmagnetic scattering due to the local mismatch in the band gap that arises when foreign atoms are incorporated into a host lattice.
We model these nonmagnetic local potentials similarly to the exchange interaction but without the possibility to induce spin flips \cite{Cygorek_Influence-of_2017}, which leads to the final contributions $H_\t{nm}^\t{e}$ and $H_\t{nm}^\t{h}$ in Eq.~\eqref{eq:Hamiltonian} for electrons and holes, respectively.

In principle one could also include the scattering with phonons in the model.
However, recent investigations have shown that their effects either are negligible for resonantly excited excitons at low temperatures \cite{Ungar_Phonon-impact_2019} or require long timescales on the order of nanoseconds when a magnetic field is applied \cite{Ungar_Phonon-induced_2019}.
Here, we are interested only in the low-temperature limit and timescales of up to $100\,$ps, so phonons can be disregarded.
For explicit expressions as well as a more detailed discussion of each constituting part of the Hamiltonian given by Eq.~\eqref{eq:Hamiltonian}, the reader is referred to Ref.~\onlinecite{Ungar_Quantum-kinetic_2017}.

In DMSs described by Eq.~\eqref{eq:Hamiltonian}, the energetically lowest exciton state consists of an electron in the conduction band and a heavy hole in the topmost valence band \cite{Bastard_Wave-mechanics_1996, Winkler_Spin-Orbit_2003}.
Whereas the electrons are characterized by a spin quantum number $s_z = \pm\frac{1}{2}$, the hh spins consist of states with angular momentum quantum number $j_z = \pm\frac{3}{2}$.
States with $j_z = \pm\frac{1}{2}$, the so-called light holes (lh), are located energetically below the hh states by the amount of the hh-lh splitting.
This splitting is a direct result of the confinement in the quantum well but is also influenced by strain \cite{Winkler_Spin-Orbit_2003}.
Focusing on systems where this splitting is large and using an excitation with $\sigma^-$ polarization, the optically prepared hh spin with $j_z = -\frac{3}{2}$ remains effectively pinned along the growth direction of the quantum well in its initially prepared state \cite{Uenoyama_Hole-relaxation_1990, Ferreira_Spin-flip_1991, Bastard_Spin-flip_1992, Crooker_Optical-spin_1997}.
Then, the description of the exciton spin dynamics can be limited to only two spin orientations for each exciton parabola, i.e., one where the exciton-bound electron spin is oriented parallel with respect to the growth direction and the exciton is bright ($s_z = \frac{1}{2}$) and another where it is flipped and thus optically dark ($s_z = -\frac{1}{2}$).
Other than states that are not coupled to the light field because of their finite center-of-mass momenta, the latter states are dark due to spin selection rules.
Denoting the two spin states by the symbols $\up$ and $\down$, respectively, the time evolution of the spin-dependent exciton density is given by
\begin{widetext}
\begin{align}
\label{eq:EoM-QKT}
\ddt n_{x_1 K_1}^\ud =&\; 
	\frac{2}{\hbar} E(t) M_\ud \Im\big[y_{x_1}^\uparrow \phi_{x_1} \big] \delta_{K_1,0}
	\pm \frac{\Jpd \nMn}{\hbar V} \sum_{x' K'} \Big( \Im\big[ Q_{\etae 3 x' K'}^{\ph{\etae} 3 x_1 K_1}\big] \pm \frac{1}{2}\Im\big[ Q_{\etae 3 x' K'}^{\ph{\etae} 0 x_1 K_1}\big] \Big)
	\nn
	& \pm \frac{\Jsd \nMn}{\hbar V} \sum_{x' K'} \Big( \sum_{ij} \epsilon_{ij3} \Re\big[Q_{-\etah i x' K'}^{\ph{-\etah} j x_1 K_1}\big] - \frac{1}{2} \Im\big[Q_{-\etah 3 x' K'}^{\ph{-\etah} 0 x_1 K_1}\big] \mp \sum_i \Im\big[ Q_{-\etah i x' K'}^{\ph{-\etah} i x_1 K_1}\big] \Big)
	\nn
	& \mp \frac{J_0^\t{e} \nMn}{\hbar V} \sum_{x' K'} \Big( 2\Im\big[ Z_{-\etah \ph{3} x' K'}^{\ph{-\etah} 3 x_1 K_1} \big] \pm \Im\big[ Z_{-\etah \ph{0} x' K'}^{\ph{-\etah} 0 x_1 K_1} \big] \Big)
	\mp \frac{J_0^\t{h} \nMn}{\hbar V} \sum_{x' K'} \Big( 2\Im\big[ Z_{\etae \ph{3} x' K'}^{\ph{\etae} 3 x_1 K_1} \big] \pm \Im\big[ Z_{\etae \ph{0} x' K'}^{\ph{\etae} 0 x_1 K_1} \big] \Big)
\end{align}
\end{widetext}
according to the QKT developed in Ref.~\onlinecite{Ungar_Quantum-kinetic_2017}.
Apart from the total exciton density $n_{x_1 K_1} = n_{x_1 K_1}^\up + n_{x_1 K_1}^\down$ in the state $x_1$ with center-of-mass wave number $K_1$, Eq.~\eqref{eq:EoM-QKT} also yields the corresponding $z$ component of the exciton spin $s_{x_1 K_1}^z = \frac{1}{2}(n_{x_1 K_1}^\up - n_{x_1 K_1}^\down)$.
For an external magnetic field oriented along the growth direction, the $z$ component is the only one relevant as both in-plane components are not occupied during the dynamics.
Note that our notation is such that each exciton state thus consists of two spin orientations, which are degenerate at zero magnetic field and show a Zeeman splitting otherwise.

In the above equation, $E(t) M_\ud$ denotes the product of the Gaussian laser pulse $E(t) = E_0\exp(-\frac{t^2}{2\sigma^2})$ with amplitude $E_0$ and width $\sigma$, with the dipole matrix element $M_\ud$ containing the optical selection rules.
The constant $\phi_{x_1} = R_{x_1}(r = 0)$ is the radial part of the $x_1$ exciton wave function evaluated at the origin, which, together with the factor $\delta_{K_1,0}$, is a consequence of the dipole approximation.
The variable that directly describes the interband transition is the optical coherence $y_{x_1}^\up$, where the spin index indicates that only the $\up$ state is optically active.
In the QKT, the dynamics is a consequence of exciton-impurity correlations $Q$ and $Z$, for which separate equations of motion must be solved.
The indices of the coupling constants $J$ in front of these correlations indicate their respective origin in terms of the Hamiltonian given by Eq.~\eqref{eq:Hamiltonian}.
Additionally, $\nMn$ denotes the impurity density in the system with volume $V$, and the summation indices $i,j \in \{1,2,3\}$ reflect the spatial directions.
We do not provide an explicit equation of motion for the impurity spin density matrix and instead assume that it is well approximated by its thermal equilibrium value, which is justified when the Mn concentration is much higher than the carrier density \cite{Thurn_Non-Markovian_2013}.
We thus also assume that the magnetic moments of the Mn impurities have had sufficient time to adjust their orientation in the applied external magnetic field before the laser pulse starts to excite any excitons.

Since the equations for the coherence as well as the various correlations are lengthy and not very transparent, we do not explicitly write them here but refer the interested reader to Ref.~\onlinecite{Ungar_Quantum-kinetic_2017}, where they were originally derived and discussed in detail.
Here, it suffices to stress that any occupation of an exciton state with finite $K$ should be viewed as a result of correlations in the system, as can be seen in Eq.~\eqref{eq:EoM-QKT}.
In contrast to a mean-field theory where the impurities are described as a homogeneous bath, the correlations in the QKT capture the breaking of the translational invariance due to the positional disorder of the impurities, which in turn is reflected by momentum nonconservation.
Thus, in the QKT, the effective independent-particle exciton states are no longer the proper eigenstates of the system \cite{Ungar_Many-body_2018}.
Rather, the proper eigenstates are determined by the equations of motion for the correlations, which include the energy-time uncertainty as well as a possible violation of strict energy conservation on the independent-particle level.

\subsection{Markov limit}
\label{subsec:Markov-limit}

To reveal the impact of quantum kinetic effects on the exciton dynamics, it is helpful to compare the results of the QKT to those of an effective independent-particle theory where all correlation effects are discarded.
Such a Markovian description can be obtained by formally integrating the equations of motion for the correlations, which have the general form \cite{Ungar_Quantum-kinetic_2017}
\begin{align}
\label{eq:Q-general}
\ddt Q(t) = i\omega Q(t) + b(t)
\end{align}
with a frequency $\omega$ that, among other contributions, contains the exciton frequencies and a source term $b(t)$ that, in general, depends on the exciton density as well as the exciton spin.
The solution to Eq.~\eqref{eq:Q-general} is given by
\begin{align}
Q(t) = \int_0^t d\tau \, e^{-i\omega(t-\tau)}b(\tau),
\end{align}
where an explicit memory appears.
Assuming $b(\tau)$ does not vary strongly within the memory time so it can be replaced by $b(t)$ and thus drawn out of the integral, the remaining integration can be solved in the limit $t \to \infty$ using the Sokhotsky-Plemelj formula \cite{Ungar_Quantum-kinetic_2017}, so the time dependence of the integral vanishes.
Physically, this also implies that one forces the system to occupy only independent-particle eigenstates (see also the Appendix in Ref.~\onlinecite{Siantidis_Dynamics-of_2001} for an extended discussion of the implications of the Markov limit).
This is in contrast to Eq.~\eqref{eq:EoM-QKT}, where the final energy eigenstates are determined by the correlations \cite{Ungar_Many-body_2018}.

Applying this scheme to Eq.~\eqref{eq:EoM-QKT} and, for numerical reasons that become clear later on, switching to a representation in frequency space, the equations of motion in the Markov limit read \cite{Ungar_Quantum-kinetic_2017}
\begin{widetext}
\begin{align}
\label{eq:EoM-MT}
\ddt n_{x_1 \omega_1}^\ud =&\; 
	\Gamma_{x_1,\omega_1} +  \frac{I M \nMn}{2\hbar^3 d} \sum_{x' \omega'} \bigg\{ \delta\big({\omega'}_{x'} - {\omega_1}_{x_1}\big) \Big(n_{x' \omega'}^\ud - n_{x_1 \omega_1}^\ud\Big) \Big[ \big( \Jsd^2 b^\parallel \pm 2\Jsd J_0^\t{e} b^0 + 2{J_0^\t{e}}^2 \big) F_{\etah x' x_1}^{\etah \omega' \omega_1}
	\nn
	& + \big( \Jpd^2 b^\parallel - 2\Jpd J_0^\t{h} b^0 + 2{J_0^\t{h}}^2 \big) F_{\etae x' x_1}^{\etae \omega' \omega_1}
	+ \big( 4J_0^\t{e}J_0^\t{h} - 2\Jpd J_0^\t{e} b^0 \pm 2\Jsd J_0^\t{h} b^0 \mp 2\Jsd\Jpd b^\parallel \big) F_{-\etah x' x_1}^{\ph{-} \etae \omega' \omega_1} \Big]
	\nn
	& + \delta\big({\omega'}_{x'} - \big({\omega_1}_{x_1} \pm \omsf\big) \big) \Jsd^2 F_{x' x_1}^{\omega' \omega_1} \Big(b^\pm n_{x' \omega'}^\du - b^\mp n_{x_1 \omega_1}^\ud\Big) \bigg\}.
\end{align}
\end{widetext}
Here, the optical excitation is subsumed in an optical generation rate of excitons given by
\begin{align}
\label{eq:optical-generation-rate}
\Gamma_{x,\omega} &= \frac{1}{\hbar^2} E(t) E_0 |M_\ud|^2 |\phi_{x}|^2 \int_{-\infty}^t d\tau e^{-\frac{\tau^2}{2\sigma^2}} \, \delta_{\omega,0}^b \delta_{x,1s}.
\end{align}
The function $\delta_{\omega,0}^b = \exp(-(\hbar\omega/2w_\t{b})^2)$ with a small value of $w_\t{b} = 1\,$\textmu eV is used to achieve a numerically scalable and stable approximation of a $\delta$ function, reflecting the fact that the resonant optical excitation occurs only at the bottom of the spin-up exciton parabola ($\hbar\omega = 0$).
Furthermore, the constant $I = \frac{3}{2}$ appears due to the envelope functions of the quantum well in the approximation of infinitely high potential barriers, $M$ is the exciton mass, and $d$ denotes the width of the well.
The Mn spin enters the equation via the spin moments $b^\parallel$, $b^0$, and $b^\pm$, which, together with the exciton form factors $F_{\eta_i x_1 x_2}^{\eta_j \omega_1 \omega_2}$, can be found in the Appendix.

The $\delta$ functions appearing in Eq.~\eqref{eq:EoM-MT} are the reason for switching to the frequency domain since then their numerical evaluation is much more convenient.
Apart from the indirect influence of the magnetic field via the moments of the impurity spin, the Zeeman energies directly appear in the energy-conserving $\delta$ functions in terms of the spin-flip scattering shift
\begin{align}
\hbar\omsf = \hbar\ome^z - \hbar\omMn^z.
\end{align}
There, the Zeeman energy of the impurities $\hbar\omMn^z = \gMn \mu_B B^z$ is subtracted from the Zeeman energy of the excitons combined with the giant Zeeman shift due to the impurities $\hbar\ome^z = \gel \mu_B B^z + \Jsd\nMn\langle S^z\rangle$.
In the energy balance together with the exciton kinetic energy $\hbar\omega_{x}$ in the state $x$, this term takes the energy cost of an exciton-impurity spin flip-flop process into account.
Similar to the QKT, the energy of an $x$ exciton is measured with respect to the $1s$ hh exciton ground state with $s_z = \frac{1}{2}$ and $j_z = -\frac{3}{2}$.

\section{Numerical simulations}
\label{sec:Numerical-simulations}

In this section, numerical simulations are performed for the QKT as well as the MT and the respective results are compared to extract the fingerprint of quantum kinetic effects in the dynamics.
For all simulations, a Zn$_{0.975}$Mn$_{0.025}$Se quantum well with a width of $20\,$nm at a temperature of $1\,$K is considered.
The optical excitation is always chosen to be resonant with the Zeeman-shifted $1s$ exciton ground state, which is excited using a Gaussian laser pulse with a full width at half maximum of $100\,$fs.
In principle, Eqs.~\eqref{eq:EoM-QKT} and \eqref{eq:EoM-MT} are valid for an arbitrary number of states.
However, here we limit the description to the four energetically lowest exciton states, i.e., the $1s$, $2s$, $2p_x$, and $2p_y$ states.
Numerically, this requires the discretization of the continuous center-of-mass momenta for each of the four states as well as their two possible spin orientations, as described in Sec.~\ref{subsec:Quantum-kinetic-model}.
The calculated exciton binding energies based on a diagonalization of the exciton problem in real space for standard ZnSe parameters (cf. Ref.~\onlinecite{Ungar_Quantum-kinetic_2017}) can be found in Table~\ref{tab:exciton-energies}.
The calculated values are in good agreement with experimental data  \cite{Cingolani_Excitonic-properties_1998, Astakhov_Binding-energy_2002, Wagner_Biexciton-Binding_2002}.

\begin{table}
	\centering
	\begin{tabular}{ccccc}
	\hline\hline
	Exciton state & \hspace{0.5cm} & Energy (meV) & \hspace{0.5cm} & $B_\t{c}$ (T)
	\\\hline
	$1s$ & & $-20.37$ & & $0.00$\\
	$2p_x$ & & $-7.14$ & & $0.83$\\
	$2p_y$ & & $-7.14$ & & $0.83$\\
	$2s$ & & $-5.35$ & & $1.31$\\
	\hline\hline
	\end{tabular}
	\caption{Calculated energies of the first four exciton states in a $20$-nm-wide Zn$_{0.975}$Mn$_{0.025}$Se quantum well measured with respect to the band gap.
	The value of the magnetic field $B_\t{c}$ indicates the threshold when the spin-flip scattering shift becomes large enough to enable a spin flip from the $1s$ state to the current state.}
	\label{tab:exciton-energies}
\end{table}

From the exciton energies one can see that the two degenerate $2p$ states lie energetically below the $2s$ state, which is a consequence of the confinement due to the quantum well in combination with the finite angular momentum quantum number of the $p$ states.
This is similar to the case of monolayer TMDs, where the $2p$ excitons are also more strongly bound than the $2s$ excitons \cite{Brem_Exciton-Relaxation_2018}.
There are several possibilities to involve excited exciton states in the dynamics.
Here, we choose the application of an external magnetic field and exploit the giant Zeeman shift of DMSs to bring the $1s$ state with a spin-up exciton-bound electron close to an excited state with a spin-down exciton-bound electron.
The necessary values of the magnetic field for such a transition are also given in Table~\ref{tab:exciton-energies} for the $2p$ and the $2s$ states.
Considering, e.g., the energy difference $E_{1s-2p} = 13.23\,$meV, a magnetic field of about $0.83\,$T is required to shift the two bands such that spin flips between them can be resonantly mediated by the exciton-impurity exchange interaction.
Thus, in the MT with strict energy conservation, one can expect that higher exciton states will become occupied if the magnetic field exceeds this value but will remain completely unoccupied for magnetic fields with a smaller magnitude.
Considering that excitons with higher principal quantum numbers are also energetically farther away from the $1s$ state, limiting the description to the four states shown in Table~\ref{tab:exciton-energies} is a good approximation as long as the magnetic field stays well below the threshold to the higher states.

\begin{figure*}
	\centering
	\includegraphics{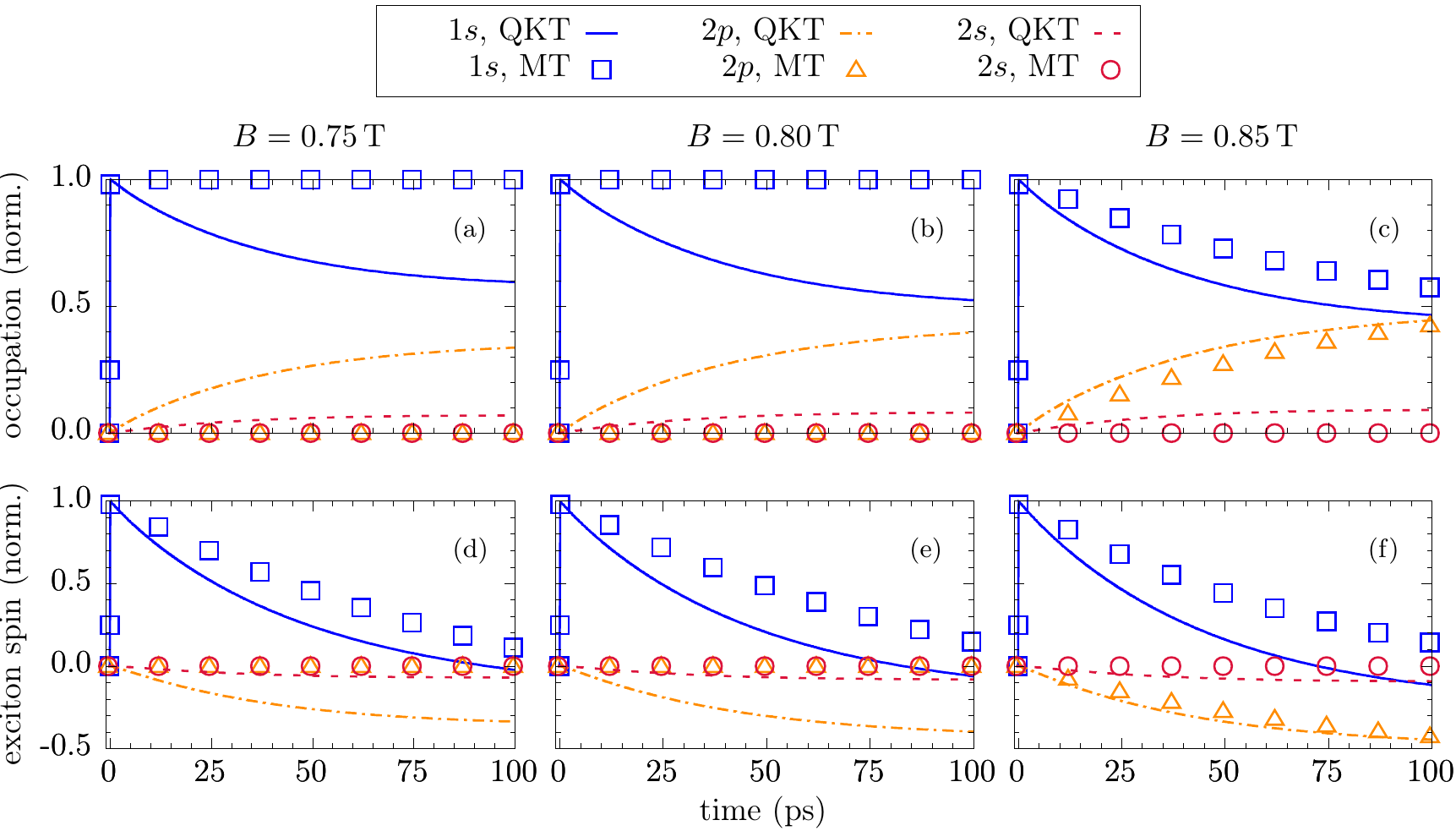}
	\caption{Time evolution of (a)-(c) the occupation of the energetically lowest four exciton states and (d)-(f) their respective spin components in the $z$ direction for three different choices of the external magnetic field $B$, as indicated.
	The occupation and the exciton spin are normalized with respect to their maximum value after the pulse, and the simulations are performed for a $20$-nm-wide Zn$_{0.975}$Mn$_{0.025}$Se quantum well excited at the $1s$ exciton resonance.
	We account for the $1s$, $2p_x$, $2p_y$, and $2s$ exciton state, where the label $2p$ denotes the sum of the degenerate $2p_x$ and $2p_y$ states.
	Results obtained by the quantum kinetic theory (QKT) are compared with those obtained by a standard Markovian theory (MT) without any memory.}
	\label{fig:occup_and_spin}
\end{figure*}

The occupation of the energetically lowest four exciton states as a function of time is plotted in Figs.~\ref{fig:occup_and_spin}(a)-\ref{fig:occup_and_spin}(c).
Note that the data are normalized with respect to the maximum occupation on the $1s$ exciton parabola reached due to the laser excitation.
For three different magnitudes of the external magnetic field, namely, $0.75$, $0.80$, and $0.85\,$T, results of simulations using the QKT and the MT are shown.
Since the two $2p$ states are degenerate, we plot only the sum of the result for the $2p_x$ and $2p_y$ states here and refer to them as the $2p$ state in the following.
Based on the Markovian model given by Eq.~\eqref{eq:EoM-MT} and the values in Table~\ref{tab:exciton-energies}, the $2p$ state should become populated only for the largest magnetic field, and the $2s$ state should remain empty for all considered magnetic fields.
Surprisingly, when comparing the results of the QKT with those of the MT, we find that the QKT predicts a sizable occupation of the $2p$ state already for the smallest chosen magnetic field and even predicts a small but visible occupation of the $2s$ state.
At $0.80\,$T, which is already close to but still below the magnetic field $B_\t{c}$ required to enable a transition in the MT, the difference between the predictions of the QKT and the MT is almost as large as $50\%$ at $100\,$ps after the pulse.
Only when the $1s$-$2p$ transition is also allowed in the MT [see Fig.~\ref{fig:occup_and_spin}(c)] do the two theories predict similar occupations for the $1s$ and $2p$ states, but deviations are still visible, and the $2s$ state remains completely empty in the MT.

The reason for the pronounced deviations between the predictions of the QKT and the MT lies in the fact that correlations are captured only in the former theory, whereas the latter is an effective independent-particle theory for excitons.
As pointed out in previous works on DMSs \cite{Thurn_Non-Markovian_2013, Cygorek_Non-Markovian_2015}, carrier-impurity correlations can cause pronounced non-Markovian effects, especially in the exciton regime due to their large effective mass as well as the proximity of optically generated excitons to the bottom of the exciton parabola \cite{Ungar_Quantum-kinetic_2017, Ungar_Trend-reversal_2018, Ungar_Many-body_2018}.
Indeed, exciton-impurity correlations are responsible for an occupation of higher exciton states when the Zeeman shift is not yet large enough to bring them into resonance with the exciton ground state.
Qualitatively, this means that the exchange interaction between excitons and impurities causes the formation of stable connections between the independent-particle energy eigenstates obtained in the MT.
These connections remain stable even at long times (see Fig.~\ref{fig:occup_and_spin}) and change the energetic structure of the problem.
Since the correlation energy amounts to several meV for the parameters considered here \cite{Ungar_Trend-reversal_2018, Ungar_Many-body_2018}, this energy can be used to overcome the energy barrier between the exciton ground state and higher exciton states as the negative correlation energy increases the exciton kinetic energy \cite{Ungar_Quantum-kinetic_2017}.
The fact that the $2s$ state becomes occupied already at Zeeman shifts comparable to the energy difference between the $1s$ and $2p$ states means that the correlation energy is roughly on the order of the $2s$-$2p$ energy splitting here.

Looking at Fig.s~\ref{fig:occup_and_spin}(a)-\ref{fig:occup_and_spin}(c), we see that the $2p$ occupation is significantly higher than that of the $2s$ state.
In part, this is because the magnetic field is chosen such that the Zeeman shift is close to or larger than $E_{1s-2p}$, that is, the $2p$ state can also be occupied without the need for correlation energy as soon as the field is sufficiently large.
More important, however, is the fact that, to reach the $2s$ state, excitons need to increase their kinetic energy by an additional amount of $E_{2p-2s} \approx 2\,$meV.
But since the scattering rate to higher energies is proportional to the exciton form factor, which quickly falls off for larger center-of-mass momenta \cite{Ungar_Phonon-impact_2019}, this process is significantly less likely.

Note that the energy-time uncertainty does not explain the observations in Fig.~\ref{fig:occup_and_spin} as it affects only the dynamics on short timescales and thus cannot explain the occupation of excited states long after the pulse is switched off.
In fact, Fig.~\ref{fig:occup_and_spin} shows that the $1s$ occupation continuously decreases throughout the whole time interval considered.
Conversely, the occupation of higher exciton states, especially the $2p$ state, continuously rises.
Using this scheme thus allows one to achieve an efficient indirect preparation of the optically dark $2p$ state with an occupation close to $50\%$ of the originally prepared $1s$ excitons on a timescale of $100\,$ps after the pulse.

\begin{figure}
	\centering
	\includegraphics{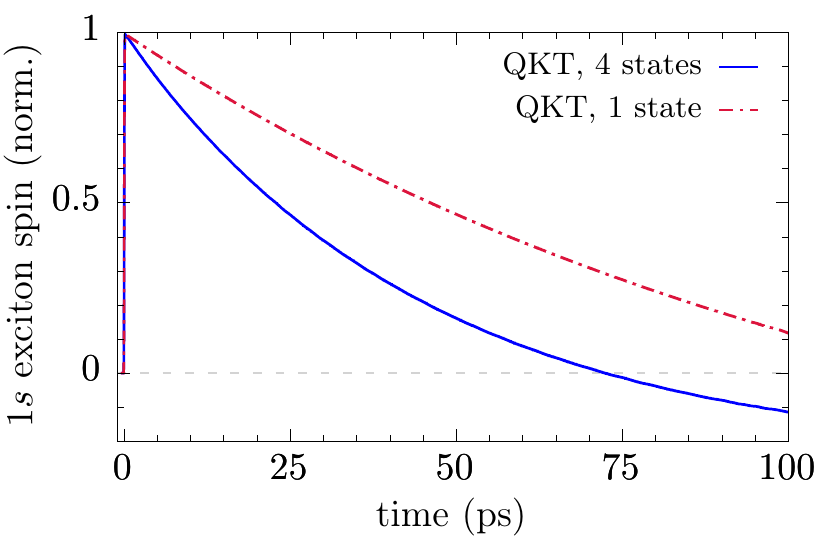}
	\caption{Time evolution of the $z$ component of the $1s$ exciton spin as obtained by the quantum kinetic theory (QKT) when the four energetically lowest exciton states (4 states) or only the exciton ground state (1 state) are accounted for.
	The simulations are performed for a $20$-nm-wide Zn$_{0.975}$Mn$_{0.025}$Se quantum well excited at the $1s$ exciton resonance in an external magnetic field $B = 0.85\,$T, and the results are normalized with respect to the maximum spin polarization after the pulse.}
	\label{fig:spin_comp}
\end{figure}

Apart from their effect on the occupation of the exciton ground state, higher exciton states also affect the spin dynamics, as can be seen in Figs.~\ref{fig:occup_and_spin}(d)-\ref{fig:occup_and_spin}(f).
As before, the results are normalized with respect to the maximum spin of the $1s$ exciton reached due to the optical excitation.
There, the occupation of higher exciton states manifests in a faster decay of the $1s$ exciton spin compared with the predictions of the MT.
This can be straightforwardly understood by thinking about the higher exciton states in terms of the possibility to open up another channel to which the spin can be transferred.
The availability of an additional channel thus causes a faster decay of the originally excited spin component.
Furthermore, Figs.~\ref{fig:occup_and_spin}(d)-\ref{fig:occup_and_spin}(f) show that the spin polarization is almost completely transferred to the $2p$ excitons on the timescale investigated here and the $1s$ spin is essentially zero after $100\,$ps, even though the $1s$ state remains occupied by about $50\%$ of the initial occupation [see Figs.~\ref{fig:occup_and_spin}(a)-\ref{fig:occup_and_spin}(c)].
Note the reversed sign of the $2p$ spin polarization compared with the optically prepared $1s$ spin, which is a consequence of the fact that $2p$ states can be reached only via a spin-flip process.

In contrast to the total exciton occupation, which remains constant after the pulse is switched off, the spin decays due to the exciton-impurity exchange interaction until it reaches a stationary value that is antiparallel with respect to the external magnetic field \cite{Ungar_Trend-reversal_2018}.
Just like for the occupations, the QKT predicts a pronounced influence of higher exciton states already below the critical magnetic field that is necessary to overcome the $1s$-$2p$ splitting on the mean-field level.
To provide a better understanding of the influence of excited exciton states on the spin dynamics, Fig.~\ref{fig:spin_comp} depicts the results of a simulation using the QKT with either only one or four exciton states accounted for at a magnetic field of $0.85\,$T.

Figure~\ref{fig:spin_comp} confirms the previous observation that the presence of higher exciton states and thus additional decay channels causes the spin to decay faster.
Looking at Fig.~\ref{fig:occup_and_spin}, this already occurs at magnetic fields that cause Zeeman shifts smaller than the separation between the exciton ground state and the lowest excited state.
Furthermore, the faster decay is already visible on short timescales of a few picoseconds and causes the spin polarization to switch its sign much sooner.
This analysis suggests that theoretical works should include transitions to higher exciton states even though they may not yet be fully resonant based on energy considerations on the independent-particle level.

\begin{figure}
	\centering
	\includegraphics{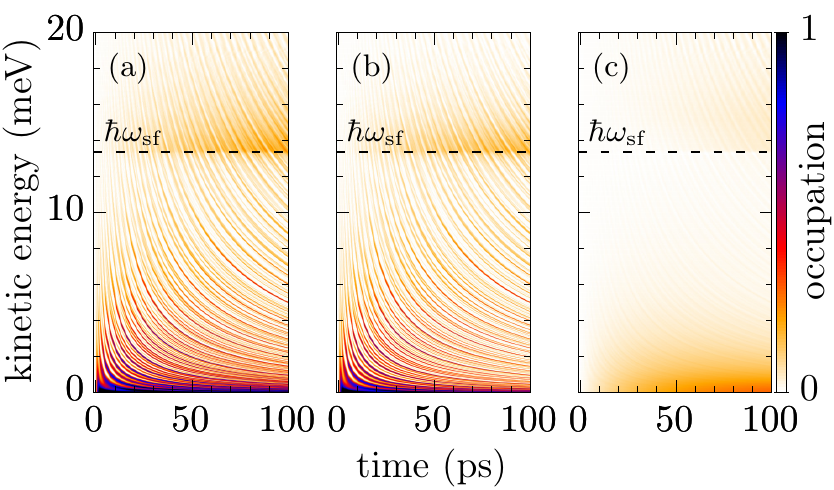}
	\caption{Time- and energy-resolved occupation of the $1s$ exciton ground state as obtained by the quantum kinetic theory when (a) only the exciton ground state or (b) the four energetically lowest exciton states are accounted for.
	The simulations are performed for a $20$-nm-wide Zn$_{0.975}$Mn$_{0.025}$Se quantum well excited at the $1s$ exciton resonance in an external magnetic field $B = 0.85\,$T. 
	The dashed line indicates the spin-flip scattering shift $\hbar\omsf$. (c) The difference between the occupations of (a) and (b).}
	\label{fig:occup_1s_diff}
\end{figure}

Coming back to the correlation energy, its impact becomes most apparent when looking at the energy- and time-resolved exciton occupation.
To this end, Fig.~\ref{fig:occup_1s_diff} displays the occupation of the exciton ground state predicted by the QKT as a function of time and kinetic energy with respect to the bottom of the $1s$ exciton parabola for a magnetic field of $0.85\,$T.
The influence of excited exciton states becomes apparent when comparing Figs.~\ref{fig:occup_1s_diff}(a) and \ref{fig:occup_1s_diff}(b) as the former takes only the exciton ground state into account, whereas the latter includes the four energetically lowest states in the calculation.
To facilitate the comparison between the two situations, Fig.~\ref{fig:occup_1s_diff}(c) displays the difference between the occupations in the two cases.

Without any correlations and in the absence of redistribution mechanisms such as phonon scattering, Eq.~\eqref{eq:EoM-MT} predicts a scattering between $\hbar\omega = 0$ and $\hbar\omsf$.
In Fig.~\ref{fig:occup_1s_diff}, these two points are represented by the bottom of the figure and the dashed line, respectively.
Thus, in the MT, one would expect only a transfer of occupations between these two points.
In contrast to that expectation, the exciton-impurity correlations captured by the QKT cause a significant occupation of states away from $\hbar\omega = 0$ that are not accessible in the MT.
Although a large fraction of excitons is still located close to $\hbar\omega = 0$ and $\hbar\omsf$, as can be seen in Fig.~\ref{fig:occup_1s_diff}, the occupation is strongly smeared out due to the correlations, which is the most prominent effect in the figure.
Note also that the redistribution of excitons towards higher kinetic energies takes place on a timescale of only a few picoseconds after the pulse, which is why some excitons quickly
gain enough kinetic energy to populate the $2s$ or $2p$ states as found in Fig.~\ref{fig:occup_and_spin}.

When comparing Fig.~\ref{fig:occup_1s_diff}(a) with Fig.~\ref{fig:occup_1s_diff}(b), the effect of higher exciton states is small but most noticeable near $\hbar\omega = 0$ as well as above $\hbar\omsf$ [see also Fig.~\ref{fig:occup_1s_diff}(c)].
There, the occupation is visibly smaller in the case when higher exciton states are accounted for since excitons are likely to get scattered to other states.
This manifests in a slightly darker region above $\hbar\omega = 0$ and $\hbar\omsf$ in Fig.~\ref{fig:occup_1s_diff}(c), that is, the exciton occupation is larger in these regions when only the exciton ground state is accounted for.
The reason for this is that, while near $\hbar\omega = 0$ the scattering to the $2p$ state is likely to occur, states about $2\,$meV above $\hbar\omsf$ possess enough kinetic energy to populate the $2s$ state and are thus missing from the $1s$ parabola.

\section{Conclusion}
\label{sec:Conclusion}

We have investigated the role of excited states in the exciton dynamics in DMS quantum wells using a quantum kinetic theory that explicitly takes correlations between excitons and magnetic impurities into account.
To enable spin-flip transitions between the exciton ground state and excited states, we have focused on systems with a sufficiently large magnetic field applied along the growth direction of the quantum well so the resulting Zeeman shift can be used to overcome the splitting between the states.
DMSs are particularly well suited for this investigation since one can exploit the giant Zeeman shift found in these materials.

By means of a comparison with a corresponding Markovian theory that can be obtained from the QKT in the limit of vanishing memory, we find that the QKT predicts a significant occupation of higher exciton states already well below the critical magnetic field that is required to bring the exciton ground state in resonance with an excited state.
The transitions can be traced back to exciton-impurity correlations that are large enough to overcome energy differences on the order of a few meV.
Thus, a sizable occupation of the optically dark $2p$ states can be reached on timescales of tens of picoseconds.

The presence of higher exciton states also has consequences for the spin dynamics, causing a faster decay of the $1s$ exciton spin since more channels are available for a spin decay.
All in all, our findings show that a faster spin decay will occur at sufficiently high magnetic fields compared to results obtained by a standard treatment using Fermi's golden rule.
Furthermore, we show that there exists an efficient indirect mechanism to populate optically dark $2p$ excitons in DMSs by applying a magnetic field and exciting the exciton ground state.
This way, a spin transfer toward states which are protected against radiative decay can be achieved.

\section{Acknowledgment}
\label{sec:Acknowledgment}

We gratefully acknowledge the financial support of the Deutsche Forschungsgemeinschaft (DFG) through Grant No. AX17/10-1.

\section*{Appendix: Impurity spin moments and exciton form factors}
\label{sec:Appendix-Impurity-spin-moments-and-exciton-form-factors}

The moments of the impurity spin $\mbf S$ appearing in Eq.~\eqref{eq:EoM-MT} are given by
\begin{align}
b^\pm =&\; \frac{1}{2}\big(\langle \mbf S^2 - (S^z)^2 \rangle \pm \langle S^z \rangle\big),
	\\
b^\parallel =&\; \frac{1}{2} \langle (S^z)^2 \rangle,
	\\
b^0 =&\; \langle S^z \rangle.
\end{align}
The exciton form factors read \cite{Ungar_Quantum-kinetic_2017}
\begin{align}
\label{eq:form-factor}
F_{\eta_1 x_1 x_2}^{\eta_2 \omega_1 \omega_2} =&\; 2\pi \int_0^{2\pi} d\psi \int_0^\infty  dr \int_0^\infty dr' \, r r' R_{n_1}(r) R_{n_2}(r) 
	\nn
	&\times R_{n_1}(r') R_{n_2}(r') J_{l_1-l_2}\big( \eta_1 K_{12}(\psi) r \big) 
	\nn
	&\times J_{l_1-l_2}\big( \eta_2 K_{12}(\psi) r' \big)
\end{align}
with $K_{12} = |\K_1 - \K_2|$ and an average over the angle $\psi$ between $\K_1$ and $\K_2$.
The index $n_i$ denotes the principle exciton quantum number, and $l_i$ is the corresponding angular momentum quantum number of an exciton in the $x_i$ state.
Furthermore, $J_i(x)$ is the cylindrical Bessel function of order $i$, the constant $\eta_j = \frac{m_j}{M}$ with $j \in \{\t{e},\t{hh}\}$ denotes the mass ratio between the carrier and exciton effective mass, and the exciton dispersion is given by $\omega = \frac{\hbar K^2}{2M}$.

\bibliography{references}
\end{document}